\documentclass[conference]{IEEEtran}
\makeatletter
\def\ps@headings{%
\def\@oddhead{\mbox{}\scriptsize\rightmark \hfil \thepage}%
\def\@evenhead{\scriptsize\thepage \hfil \leftmark\mbox{}}%
\def\@oddfoot{}%
\def\@evenfoot{}}
\makeatother
\usepackage{graphicx}
\usepackage{amssymb}
\usepackage{amsmath,url}
\usepackage{multicol,multirow}
\usepackage{epsf,psfrag,pstricks}
\usepackage{rotating,subfigure}
\usepackage{url}

\begin{document}

\title{TorrentGuard: stopping scam and malware distribution in the BitTorrent ecosystem}


\author{\IEEEauthorblockN{
Michal Kryczka\IEEEauthorrefmark{1}\IEEEauthorrefmark{2},
Ruben Cuevas\IEEEauthorrefmark{2},
Roberto Gonzalez\IEEEauthorrefmark{2}
Angel Cuevas\IEEEauthorrefmark{3} and
Arturo Azcorra\IEEEauthorrefmark{2}}

\IEEEauthorblockA{\authorrefmark{1}Institute IMDEA Networks}
\IEEEauthorblockA{\authorrefmark{2}University Carlos III Madrid}
\IEEEauthorblockA{\authorrefmark{3}Telecom SudParis}

}

\maketitle

\begin{abstract}

In this paper we conduct a large scale measurement study in order to analyse the fake content publishing phenomenon in the BitTorrent Ecosystem.  Our results reveal that fake content represents an important portion (35\%) of those files shared in BitTorrent and just a few tens of users are responsible for 90\% of this content. Furthermore, more than 99\% of the analysed fake files are linked to either malware or scam websites. This creates a serious threat for the BitTorrent ecosystem. To address this issue, we present a new  tool named TorrentGuard for the early detection of fake content. Based on our evaluation this tool may prevent end users from downloading more than 35 millions of fake files per year. This could help to reduce the number of computer infections and scams suffered by BitTorrent users. TorrentGuard is already available and it can be accessed through both a webpage or a Vuze plugin.



\end{abstract}

\section{Introduction}


BitTorrent is one of the most popular applications in the current Internet. It is daily utilised by millions of users and is responsible for a major portion of the Internet traffic \cite{Sandvine}. This success motivated the research community to investigate different aspects of BitTorrent covering performance \cite{NikosBitMax}\cite{BitTyrant}, economics \cite{ONO}\cite{INFOCOM}\cite{P4P} and incentives \cite{IzhakInfocom}\cite{FairTorrent} issues. However, to the best of the author knowledge, the research community has put less attention to BitTorrent security aspects. Some previous works have analysed the vulnerabilities of the BitTorrent protocol to free-riders \cite{LiogkasFreeRiders}\cite{LocherFreeRiders}\cite{SirvianosFreeRiders} whereas some others  address the lack of privacy offered by BitTorrent \cite{Legout10}. More recently, in a previous work \cite{CONEXT} we demonstrated that the BitTorrent ecosystem is suffering from a continuous \emph{poisoning index} attack resulting in 30\% of published torrents associated to fake content. Furthermore, this fake content produces 25\% of the download events, which means that every fourth content download in BitTorrent is fake. These initial results highlight a serious issue that, to the best of the authors knowledge, has still not been covered by the research community.

In this paper we thoroughly analyse the \textit{fake publishing} phenomenon in BitTorrent in order to understand its real impact on the system performance as well as the potential risks of fake content for BitTorrent users. Furthermore, we propose a practical solution to mitigate this problem. We base our study on data collected from torrents published in The Pirate Bay portal during a period of 14 days from 30-04-2011 to 13-05-2011. The 35\% of almost 30K analysed torrents are associated to fake content. This depicts a 5\% increment in the presence of fake content within the BitTorrent ecosystem in a period of one year between our two measurement studies. This justifies (even more) the necessity of the research conducted in this paper.

In order to fight the fake publishing phenomenon, the first step is to properly characterise the fake publishers and their behaviour. The current BitTorrent portals solutions identify fake publishers through the user account that they use to upload fake torrents to the portal. We show in the paper that this technique is inefficient since the fake publisher can generate as many user accounts as needed in those portals. Instead, the parameter that uniquely identifies the fake publisher is the IP address it uses to perform its activity. Surprisingly, our data reveals that just 20 fake publishers (whose IP we identify) are responsible for injecting 90\% of fake content in the BitTorrent ecosystem. Moreover, most of these IP addresses belong to Hosting Providers where the fake publishers rent dedicated high-resource servers to perform their activity. 

The fake publishing activity is time consuming since a fake publisher needs to manually create the user accounts used in the different portals (in some cases up to 4 accounts per day). Furthermore, this activity requires dedicated resources (e.g. rented servers). This investment in time and resources can be only justified by a strong motivation behind the distribution of fake content. We have downloaded and manually inspected a large number of fake content published during our measurement period and found 3 different profiles among the fake publishers: $(i)$ a first group of fake publishers aims to spread malware using the popular BitTorrent system; $(ii)$  a second set of users tries to attract BitTorrent users to scam websites in order to get economical benefit from the victims by using different scam techniques; $(iii)$ the last group is formed by antipiracy agencies that upload fake versions of those content that they want to protect.

Our data shows that more than 99\% of the published fake content is associated with the two first profiles. This supposes a very serious threat for the BitTorrent ecosystem since the activity of these publishers may lead to thousands of undesirable episodes of scammed users and computer infections. These findings suggest that new solutions need to be proposed in order to eliminate or at least reduce the number of fake content available in the BitTorrent ecosystem. Towards this end, we have designed and implemented  TorrentGuard. This is a novel detection tool that allows to identify the IP address of the fake publisher, thus being able to report as fake each content published from this IP address at the moment of its publication. Based on the performed evaluation, TorrentGuard would be able to avoid more than 35 millions fake content downloads every year. This means, preventing hundreds of thousands of users to suffer from computer infections or scam incidents every year. TorrentGuard can be currently used through a publicly available website and a Vuze plugin.

The rest of the paper is structured as follows. Section \ref{sec:background} presents the background information. In Section \ref{sec:dataset} we describe our measurement methodology and present our dataset. Next, Section \ref{sec:malicious} characterises fake publishers, while Section \ref{sec:malicious_content} classifies them depending on the goal they pursuit with their activity. Section \ref{sec:downloaders} shortly characterises the downloaders of the fake content. In Section \ref{sec:solution} we describe and evaluate our solution to improve the detection of fake content. We also discuss possible countermeasures to TorrentGuard and their efficiency. Section \ref{sec:related_work} describes relevant works to this paper. Finally, Section \ref{sec:conclusions} concludes the paper.

\section{Background}
\label{sec:background}

\begin{figure*}[t!]
\centering
\includegraphics[width=4.6in]{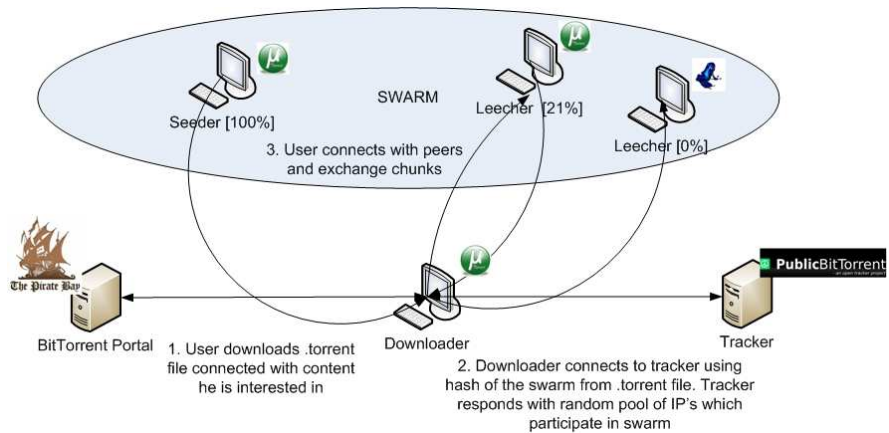}

\caption{BitTorrent ecosystem basic functionality}
\label{fig:ecosystem}

\vspace{-0.7cm}

\end{figure*}

In this Section we briefly describe the main aspects of the BitTorrent ecosystem making an special emphasis on the procedure of publishing content on The Pirate Bay (and by extension on other BitTorrent portals) and specifically, how fake publishers do it.  This is summarised in Figure \ref{fig:ecosystem}. For a full description of the BitTorrent ecosystem we refer the reader to \cite{comm} and \cite{Zhang10}.

\subsection {Main elements of BitTorrent ecosystem} 

\noindent \emph{- BitTorrent Portals:} these are webpages which index .torrent files, classify them into different categories and provide basic information for each file. These portals serve as rendez-vous points between content publishers and BitTorrent downloaders. The publishers upload their .torrent files to BitTorrent portals and the clients download them.

\noindent \emph{- .torrent file:} this is a meta-information file including relevant information for the BitTorrent protocol such as: $(i)$ the content infohash, this is a unique identifier of the content in the BitTorrent ecosystem; $(ii)$ the IP address of the BitTorrent Tracker managing the content distribution process; $(iii)$ the size of the content and the number of pieces forming the file.

\noindent \emph{- magnet link:} this is an URI-like link that includes the infohash of an specific content and optionally the address of a tracker \cite{magnet-wikipedia}. A user can launch a download process retrieving the magnet link instead of the .torrent file from  a BitTorrent portal. Then, with the magnet link the user can obtain the .torrent file from other peers in the swarm\footnote{Also the magnet link can be used as index to retrieve the associated .torrent file from the different DHTs implemented by BitTorrent clients \cite{bittorrent-dhts}.}. The magnet links have recently become significantly important since the administrators of the largest BitTorrent portal, The Pirate Bay, have announced their intention to stop serving .torrent files from March 1st 2012. Instead, they will serve exclusively magnet links \cite{magnet}.


\noindent \emph{- BitTorrent Trackers:} these are servers which manage the BitTorrent download process of a given content. The set of peers downloading a given file is named \emph{swarm}. The tracker maintains a list with the IP addresses and the download progress of all the peers forming the swarm associated to a specific content. Furthermore, when a new peer joins the swarm, it contacts the tracker in order to obtain a list of IP addresses of other peers participating in the swarm. By doing so, the new incomer is able to retrieve pieces of the content from these peers.

\noindent \emph{- BitTorrent downloaders (peers):} these are clients forming the swarm that download and/or upload pieces of the content. We distinguish two types of peers. A \emph{seeder} is a peer that possess a complete copy of the content, thus only uploads pieces whereas a \emph{leecher} does not have the complete file so that it uploads and downloads pieces.

\noindent \emph{- BitTorrent publishers:} these are the clients that make available the first copy of the content in the BitTorrent ecosystem.

\subsection{Publishing a content in BitTorrent}

When a publisher wants to publish a content in the BitTorrent ecosystem, it firstly creates a .torrent file.  After creating the .torrent file, the publisher uploads it to one or more BitTorrent portals. For this purpose, it uses a user account (with a specific username) created in these portals. Furthermore, the publisher distributes the first copy of the content by acting as the initial seeder in the associated swarm. Therefore, \emph{the content publisher can be identified by the IP address of the initial seeder distributing the content and by the username utilised to upload the content to a BitTorrent Portal}.

In this paper we specifically address the fake content publishing phenomenon in BitTorrent. A fake publisher is a user that exploits the BitTorrent ecosystem to publish fake content, this is, content that is different than what is expected from the content name. A fake publisher makes available the fake content from a single IP address (or limited number of IP addresses) that corresponds to the initial seeder of all its published content. Furthermore, a fake publisher typically creates a user account in a BitTorrent portal from  which it uploads .torrent files associated with its fake content. Some portals, such as The Pirate Bay, remove this user account after some client reports that it is being used to publish fake content. Then, the fake publisher reacts by creating a new account to publish new .torrent files and this loop keeps repeating. Hence, contrary to the case of regular publishers (that can be identified by its associated username in the BitTorrent portal), fake publishers can exclusively be identified by its IP address.
Finally, it must be noted that, to the best of the authors knowledge, the  previously described technique based on users' reports is the only one used nowadays for detecting and deleting fake content.

\subsection{Downloading a content in BitTorrent}
When a user wishes to download a content, it first downloads the .torrent file associated to the content from a BitTorrent portal such as The Pirate Bay. Then, the user retrieves the IP address of the Tracker managing the swarm from the .torrent file and connects to it. The Tracker provides the user with a list (50 to 200) of IP addresses participating in the swarm along with the number of seeders and leechers forming the swarm. Finally, the user starts downloading the pieces of the content from the obtained IP addresses.

\subsection{BitTorrent Portals, the case of The Pirate Bay}

We use The Pirate Bay as the reference BitTorrent Portal for our study. Previous works \cite{Zhang10} have demonstrated that the Pirate Bay is a key element and the most important portal in the BitTorrent ecosystem. A publisher needs to create a user account in order to upload .torrent files to The Pirate Bay whereas other portals such as IsoHunt \cite{ISOHUNT} use crawling techniques to obtain the offered content from third portals such as The Pirate Bay. 
Hence, The Pirate Bay is the most interesting portal to be considered in order to understand the content publishing phenomenon in BitTorrent. Specifically, The Pirate Bay offers the following relevant services to our study:  $(i)$ an RSS feed system in which each new published content is announced along with the username that uploaded the .torrent file to the portal; $(ii)$ each user registered within The Pirate Bay portal has an individual webpage in which its published torrents are listed and $(iii)$ The Pirate Bay removes the accounts, webpages and .torrent files of those users whose content is detected as fake. Typically, this happens after a client, who downloaded the content, reports its falseness to The Pirate Bay administrators.


\section{Measurement Methodology}
\label{sec:dataset}


This Section describes our measurement methodology to identify and characterise the main properties of the fake publishers (i.e. users publishing fake content). For this purpose we crawl The Pirate Bay, the most popular BitTorrent portal (as reported by previous works \cite{Zhang10} and by Alexa Ranking \cite{alexa}). 

The main objective of our measurement study is to identify fake publishers. Towards this end, our measurement tool has three independent modules. The first one is responsible for finding the IP address and username of the publisher associated with each announced content in The Pirate Bay. For this purpose, the module is subscribed to the RSS feed of The Pirate Bay in order to learn each torrent just after its birth. After getting a new .torrent file the tool obtains the username that uploaded the .torrent file to The Pirate Bay. Furthermore, it uses the infohash within the .torrent file\footnote{Note that we have implemented a new functionality that allows our tool to get the infohash also from magnet links.} to connect to the associated Tracker to obtain the IP addresses of the peers forming the swarm in its very initial stage. Then, it is very likely that we can find the IP address of the content publisher (initial seeder). Specifically, we face three different situations: $(i)$ The tracker only reports the IP address of the initial seeder. This is likely to happen since we connect to the swarms just after the torrent birth. $(ii)$ The tracker announces the presence of one seeder and few leechers in the swarm. Then, by connecting to all these peers and obtaining their bitfields (vector that shows the number of pieces that a peer possesses) we are able to identify which one is the initial seeder, and thus the content publisher. $(iii)$ In some cases, the Tracker announces the presence of quite a few seeders in the swarm thus we cannot identify the initial seeder. This happens because the swarm has been formed before the torrent is announced in the RSS feed of The Pirate Bay portal. Therefore, using the described methodology we are able to characterise the content publisher by both its username and IP address in many cases. 

The second module of our tool is responsible for identifying those publishers that are in fact fake ones. For this purpose our tool connects periodically (every 5 minutes) to the Pirate Bay webpage of each known publisher. If at some point the Pirate Bay webpage has been removed we consider that the IP address associated with the removed account belongs to a fake publisher. Furthermore, we also collect the time that The Pirate Bay requires to detect and eliminate each fake publisher account.

Finally, our tool has a third module that counts the number of peers that connect to the swarm of each fake content in order to download it. Specifically, our tool systematically queries the Tracker managing the download of each fake content to obtain those IP addresses participating in the swarm. In order to accelerate this process we perform this task from four independent machines.

\subsection{Dataset description}

We have applied the described methodology between 30-04-2011 and 13-05-2011, in addition to 5 days of warm-up phase dedicated to identify the initial fake publishers' IP addresses. During the measurement period we have collected 29330 torrents, from which 10206 (35\%) were identified as fake ones. Furthermore, we have collected the IP addresses of those peer participating in swarms associated with fake content until  two instants: $(i)$ the moment the content is removed from The Pirate Bay and $(ii)$ the end of our measurement study.


\section{Fake Publishers Characterization}
\label{sec:malicious}

\begin{figure}
\centering
\includegraphics[width=2.4in]{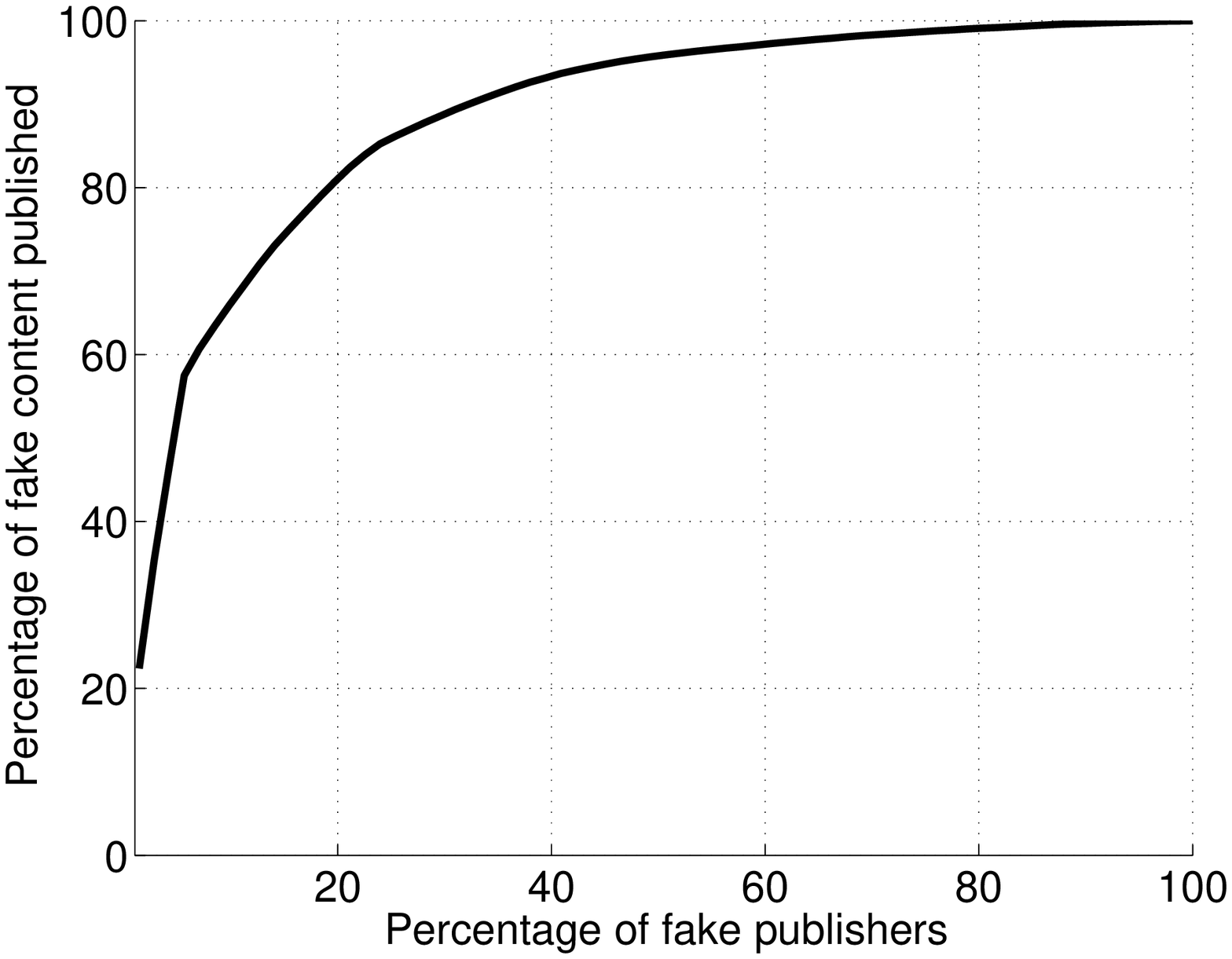}

\caption{Percentage of fake content published by the top x\% fake publishers}
\label{fig:ip_per_torrent}
\vspace{-0.6cm}

\end{figure}

Our results reveal that more than 1/3 of the content published in the Pirate Bay is fake. This shows an increasing trend in the number of fake content regarding our previous study done one year earlier when the fake content represented a 30\%. Therefore, it is critical to eliminate or at least reduce this huge number of fake content in the BitTorrent ecosystem. The first step towards this end is to identify who is responsible for publishing this fake content and characterising its behaviour. In this Section, we address this issue  using the collected data. More specifically, we aim to answer questions such as: \emph{How many fake publishers (i.e. IP addresses) are uploading fake content to the BitTorrent Ecosystem?}, \emph{From where (i.e. which ISP) they perform their activity?} or \emph{How frequently they upload fake content?}.

\subsection{Number and Contribution of Fake Publishers}

Unexpectedly, we observe that only 71 IP addresses are responsible for those 4779 fake content for which we identified the initial seeder. This implies almost 70 fake content published from each of these IPs in average. However, it is interesting to investigate the level of the contribution of each one of these fake publishers. Towards this end, Figure \ref{fig:ip_per_torrent} depicts the percentage of fake content published by the top x\% of these fake publishers. The graph shows a skewed distribution where 10 IPs (14\%) are responsible for publishing almost 75\% of all the fake contents. Moreover, this number increases to 90\%  if we consider the top 20 IP addresses (28\%). Therefore, we can conclude that a reduced number of just 20 fake publishers are responsible for poisoning the BitTorrent ecosystem. In the rest of the paper we focus on thoroughly studying this group of 20 fake publishers that we refer to as \emph{Top Fake Publishers}.

%
%
%

\subsection{Location of fake publishers}

We have mapped the IP address of each one of the Top Fake Publishers to its correspondent ISP using the MaxMind database \cite{maxmind}. Surprisingly, 17 out of the Top 20 fake publishers operate from Hosting Providers. These are companies dedicated to rent high-resources (cpu, memory and bandwidth) provisioned servers. Moreover, 70\% of the fake content is seeded from just two Hosting Providers named \emph{OVH Systems} and \emph{Obtrix} located at France and New Zealand respectively.

On the one side fake publishers need resources in order to sustain the distribution of a large number of fake files \cite{CONEXT} and on the other side anonymity due to the \emph{illegitimate} activity being performed. The usage of rented servers in Hosting Providers covers both requirements.

Hence, the use of dedicated servers in Hosting Providers reveals that most of the fake publishers perform their activity from a stable IP since those servers typically have a static IP address configured. This makes them easily identifiable. In this sense, the usage of anonymity services such as TOR \cite{tor} or proxy services seems to be useful for fake publishers in order to make difficult their identification. However, we have not found that the fake publishers identified in our dataset use such services. This suggests that the severe performance degradation associated to these anonymity services prevent fake publishers from using them. We further discuss these aspects in Section \ref{sec:countermeasures}.

\begin{figure}
\centering
\includegraphics[width=2.4in]{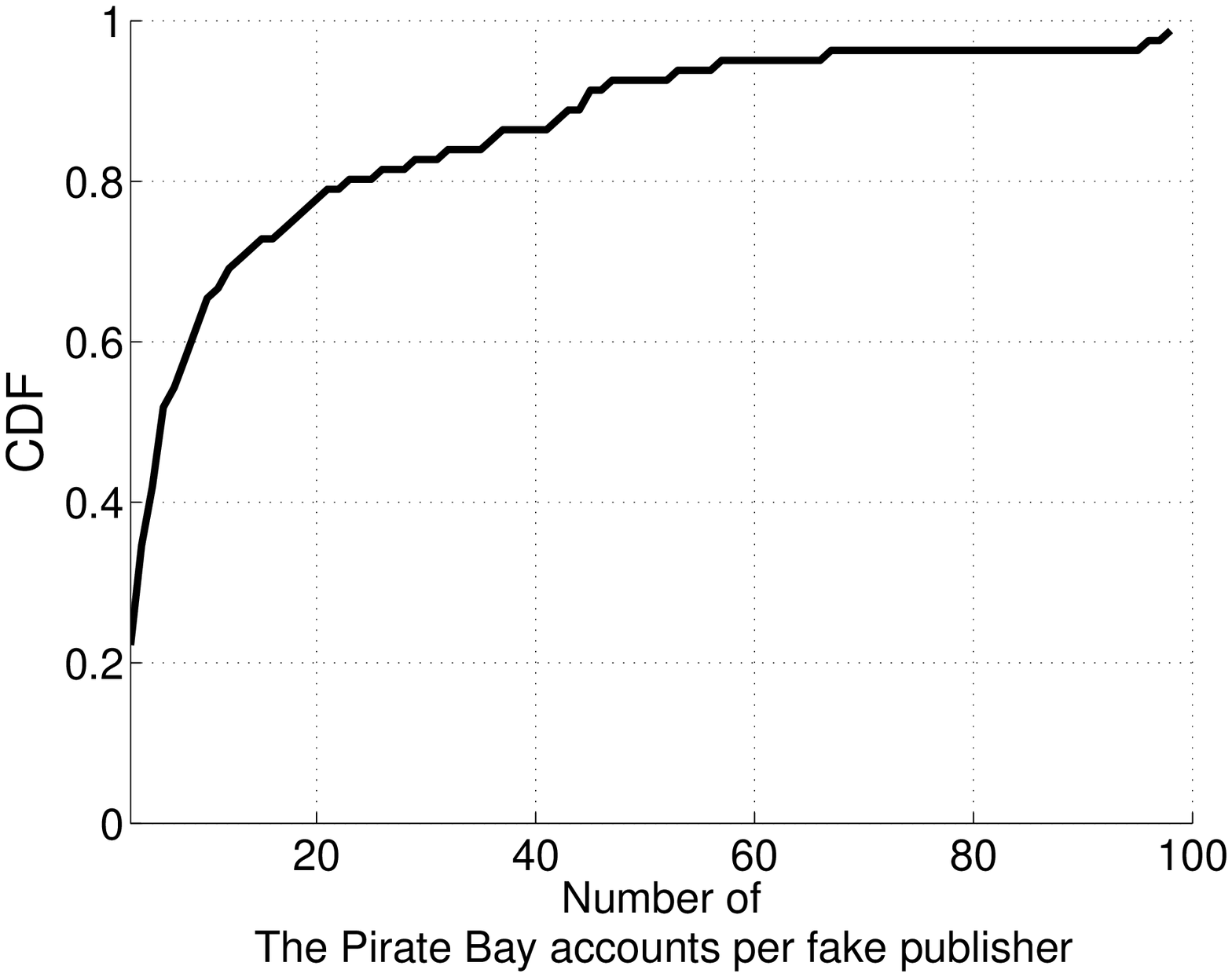}

\caption{CDF of the number of The Pirate Bay accounts per fake publisher}
\label{fig:cdf_ip_user}
\vspace{-0.6cm}

\end{figure}

%
%
%
%

\subsection{Pirate Bay accounts utilisation}

The Pirate Bay solicits to solve a CAPTCHA \cite{captcha} in order to create an account to avoid the automatic generation of accounts. Hence, fake publishers are obeyed to create their accounts manually. Figure \ref{fig:cdf_ip_user} shows the CDF of the number of The Pirate Bay accounts used by each one of the 71 identified fake publishers. A fake publisher use (in median) 6 accounts in a period of 14 days. However, a 5\% of the fake publishers inject content using more than 58 different accounts in the same period. This represents an average number of 4 accounts per day. This result suggests that fake publishers need to dedicate time to track the availability of their accounts in order to manually generate new ones if needed.

Interestingly, we also observe a second strategy that although  marginal is worth to report. In these cases, fake publishers hijack the accounts with a legitimate publishing history. This provides a trust reputation among the downloaders. Therefore, this could extend the time that fake user could be injecting fake torrents before being reported. However, due to the required technical skills for applying this technique, this case represents less than 1\% of all fake accounts.

\subsection{Publishing Strategies}

Fake users follow two different strategies to upload fake contents into The Pirate Bay portal. On the one hand, we found users that publish a large number of fake content in a row (typically around 10) in just few seconds after creating a user account. Once the account is deleted, they repeat the same process from a new account. Around 70\% of Top Fake Publishers use this technique. On the other hand, 30\% of the Top Fake Publishers upload just one or two fake contents with a username. This is a more conservative technique that extends the time that those fake accounts are active before being eliminated when compared to the previous case. Specifically, the accounts of those publishers using the first strategy are detected and then deleted in 92 minutes (in average) whereas the accounts of those using the second strategy are deleted in 253 minutes, thus being their content available 2.75 times more time in The Pirate Bay. Unexpectedly, although the second strategy offers longer accounts' lifetime, it attracts only 47 downloaders per torrent (in average) in front of the 84 attracted by fake publishers using the first strategy. This happens because the fake publishers using the first strategy typically use popular names associated to their content whereas publishers using the second more conservative strategy do not use such popular names.


\subsection{Strategies to attract downloaders}

The main goal of fake publishers in BitTorrent is to produce as many downloads of their content as possible. Therefore, they need to offer torrents that sound very attractive for the downloaders. Towards this end, we have observed that fake publishers use three different strategies: \emph{(i)} they assign to the content a very popular name such as the title of the last released Hollywood movies; \emph{(ii)} creating the false impression that the content has been published by a well-known and trusted user. For this purpose, the fake publisher names its content in the same way as the trusted one. For instance, eztv one of the most popular publisher in The Pirate Bay adds the signature [eztv] at the end of the title of its published files. Then, some fake publishers also add this signature to the title of their fake content; \emph{(iii)} presenting attractive performance statistics (i.e. a  high number of seeders and leechers) for the fake torrent. In this way, the fake torrent is  perceived as a very popular torrent by the downloaders, that assume they will obtain a high download rate in case of selecting that torrent. To generate these fake statistics the publisher connects to the Tracker many times using a single IP but different ports. The tracker considers each of these IP+port pairs as a single peer and reports a high number of seeders and leechers. The Pirate Bay retrieves and presents these statistics from the Tracker.

\vspace{0.2cm}
\emph{In summary, the fake content publishing activity is performed from Hosting Providers facilities by just few dozens of users. Furthermore, fake publishers are aware of how the BitTorrent ecosystem works, thus they use sophisticated strategies in order to improve the success of their activity.}

\section{Fake Publishers Profiles}
\label{sec:malicious_content}

After characterising the Fake Publishers behaviour, we still need to answer an important question: \emph{What incentives a user has to publish fake content?}. 
To answer this question we have downloaded up to 10 files published by each fake publishers in our dataset and manually inspected them. Our analysis reveals the presence of three different profiles: malware propagators, scammers and antipiracy agencies. Next, we describe in detail each one of these profiles.

\subsection{Malware propagators}

These users exploit the popularity of BitTorrent in order to rapidly propagate malware among thousands of users. On the one hand, for some of the users in this group the published content is the malware itself. In this case, the  content including the malware pretends to be typically a patch for a popular game, a key generator, etc. On the other hand, a second set of users use a more sophisticated technique. They publish a movie with a catchy title. The content has the standard size of a DivX movie (i.e. between 700MB and 1GB), and even sometimes includes a second small file with a real sample of the movie. Hence, the file has the appearance of a (non-fake) legitimate content. However, when a user downloads the content and tries to play the movie, it is requested to reproduce it using Windows Media Player (WMP) in case a different player is run instead. When the movie is finally reproduced with the WMP a pop-up window appears requesting to install new codecs along with an url link from where these codecs can be downloaded. Of course, the file including those pretended codecs is reported as a malware by anti-virus software.

\subsection{Scammers}

In this case, the fake publisher uses a similar technique to the sophisticated one described above. However, when the user plays the movie with WMP, it is automatically redirected to a website in the Internet. A second variant used by scammers is to provide a file protected with a password (typically .rar), and offer the user a website in which the password can be obtained. Once the user gets into one of these websites, a credit card payment is requested in order to obtain some privilege to watch the downloaded movie (e.g. the password of the .rar file). In some other situations the user is informed that in order to check he is not a bot, a survey must be filled previously to watch the movie. This survey results to be a contest in which the client is obeyed to subscribe for a paid premium SMS service. These websites are often reported as scam on different forums. An example of them is \url{http://movieyt.com}.
 
%
%
%
%
%

We have performed a more detailed analysis of these websites. On the one hand, when a user wants to abandon the webpage several pop-up windows appear trying to change user mind and making leaving the webpage at least bothersome. On the other hand, when a user enters some of these webpages, a pop-up window advertising a Facebook group of the webpage shows up. This pop-up does not react to the explorer close button, rather, just by clicking on the ``I like it'' Facebook button the window closes. This method aims to increase the trust of the webpage so that users interpret it is a legitimate website. More importantly, this finding suggests that these scammers do not limit their activity to BitTorrent but they also try to capture victims from other popular applications such as online social networks.




\begin{table}[lt!]

\center

\begin{tabular}{||p{0.1\textwidth}| p{0.1\textwidth} | p{0.1\textwidth}| p{0.1\textwidth}||}
\hline \hline Country &  Percentage of BitTorrent users downloading fake content & Percentage of BitTorrent users & The ratio\\ 
\hline \hline 
United States	&	12.40\%&10.42\% & 1.19 \\
China	&	6.27\%&4.20\% & 1.49 \\
Great Britain&	4.60\%	&6.26\% & 0.73 \\
Brazil	&	4.26\%&2.68\% & 1.59 \\
Italy	&	3.88\%&4.13\% & 0.94 \\
India	&	3.78\%&5.71\% & 0.66 \\
Canada	&	3.29\%&3.85\% & 0.85 \\
Spain	&	2.79\%&5.95\% & 0.47 \\
Austria	&	2.73\%&2.83\% &  0.96\\
Poland	&	2.66\%&2.86\% & 0.93 \\
\hline \hline 

\end{tabular} 

\caption{Demographics of BitTorrent users vs fake content downloaders per country (the third column represent the ratio column 1/column 2)}
\label{tab:downloaders_fake}
\vspace{-1.2cm}
\end{table}

\subsection{Antipiracy Agencies}

The two previous profiles have dishonest purposes. Antipiracy agencies instead, publish fake versions of the copyrighted content that they want to protect. This content however, is not what downloaders are expecting from the title (e.g. copyrighted movie). Sometimes this content includes antipiracy adverts. The action performed by antipiracy agencies is limited in the number of contents (under request from a company) and time (in the weeks before and after the content, e.g. movie, is released).

\vspace{0.2cm}
\emph{In summary, we distinguish three different profiles among fake publishers. On the one hand, 65\% of the Top Fake Publishers in our dataset are malware propagators and are responsible for around a 30\% of the published fake content. On the other hand, a 35\% of the Top Publishers are scammers and they published a 70\% of the fake content during our measurement period. Finally, antipiracy agencies represent a very small fraction of the fake content published due to the specificity of their actions. Therefore, most of the fake content published (by malware propagators and scammers) is potentially harmful, specially for not technically skilled downloaders. This represents a serious risk for the BitTorrent, and by extension for the whole Internet, that should be erased or at least mitigated. We address this issue in Section \ref{sec:solution}}.

\begin{figure}[t!]
		\centering
		\includegraphics[width=2.4in]{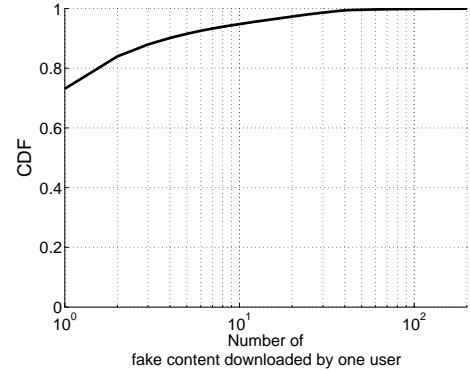}

		\caption{CDF of the number of fake content downloaded by one user}
		\label{fig:cdf_downloaders}
\vspace{-0.7cm}
\end{figure}

\section{Characterizing the Downloaders of Fake Content}
\label{sec:downloaders}

In this Section we look at the studied phenomenon from the victims side. First we analyse the demographics of the victims and group them per country in order to understand which countries suffer more from the reported problem. In order to provide full meaningful results we have compared the demographic distribution of the victims of fake content and the general demographic distribution of BitTorrent clients obtained from the dataset used in our previous work \cite{CONEXT} that includes the IP address of 27M clients associated to around 40k torrents.

Table \ref{tab:downloaders_fake} offers the obtained results. It shows the percentage of victim downloaders of fake content, the percentage of BitTorrent users and the ratio between these two percentages for the 10 countries with a larger number of victims. If the victims were randomly selected, this ratio would be close to 1. However, this is not the case. On the one hand, we observe that some countries such as US, China and Brazil show a ratio $>$ 1. For instance, Brazil has a ratio equal to 1.59. This means that Brazil has 59\% more victims than expected from a random process. On the other hand, countries such as UK, India or Spain show a value $<$ 1. For instance Spain has a ration equal to 0.47. This means, Spain only has 47\% of the victims it should have from a random process.
 
Next, we study the number of fake content downloads performed by a single user. This help to understand whether there are users that are highly vulnerable to the described threats. Figure \ref{fig:cdf_downloaders} shows the CDF of the number of fake content downloaded by a each victim. We can see that 70\% of the victims downloaded just 1 fake content. However, it is worth to note the presence of hundreds of users who downloaded multiple fake torrents during the measurement period.

\emph{In a nutshell, the obtained results suggest that users from some specific countries (those having a ratio less than 1) are more skilled to identify fake content so being more protected against possible infections and/or scam episodes. More importantly, we have revealed that hundreds of users in our dataset download more than 5 fake content in a period of two weeks. These seems to be non-skilled users that are seriously exposed to scammers and malware propagators.These highly vulnerable users are the ones that will potentially obtain a higher benefit from the system described in the next section.}


\section{TorrentGuard}
\label{sec:solution}

\begin{figure*}
\centering
\includegraphics[width=4.6in]{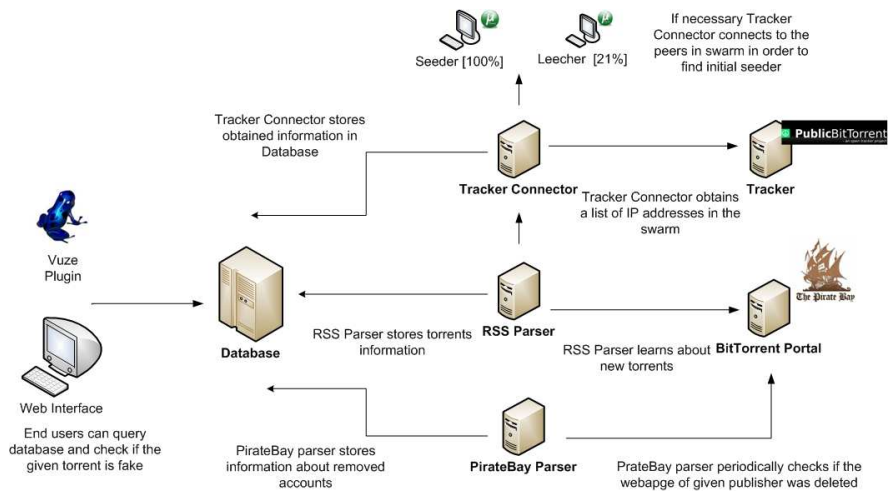}

\caption{The schema of TorrentGuard}
\label{fig:TorrentGuard}
\vspace{-0.6cm}

\end{figure*}

In the previous Sections we have demonstrated that a large number of fake content (35\%) is currently being published in the BitTorrent ecosystem, and what is worse, most of these fake content are potentially harmful for those users that download them. We have also seen that the techniques used to remove these contents are inefficient and require heavy human intervention to: first, detect and report the falseness of a given content, and second, remove it from the BitTorrent portals (this is done by the portal administrator). Furthermore, the scope of the user reports is limited to a single BitTorrent Portal, thus the content is removed exclusively from this portal instead of the whole BitTorrent ecosystem.

In this Section we present our tool, named TorrentGuard, that aims to automatise and accelerate the process of detecting fake publishers. For this purpose, TorrentGuard identifies a fake publisher by its IP address instead of its username as it is done by BitTorrent portals such The Pirate Bay nowadays. By doing so, a fake content can be identified just after its birth since we can identify that the IP address of the initial seeder belongs to a fake publisher. This allows to accelerate the detection process.

Furthermore, contrary to current techniques used by BitTorrent portals, TorrentGuard removes the fake content from the whole BitTorrent ecosystem because it reports the content infohash. Since the infohash uniquely identifies a content in the BitTorrent ecosystem, a user of TorrentGuard  can identify the content as fake independently of the portal from which the .torrent file was retrieved or even if it was obtained from the BitTorrent DHT service.

In the rest of the Section we present the details of the TorrentGuard implementation as well as the performance results obtained over a testing period of 14 days. 

\subsection{TorrentGuard Implementation}

Figure \ref{fig:TorrentGuard} depicts a complete schema of TorrentGuard. It is composed by the following modules:

\begin{itemize}
\item \emph{RSS Parser}: this module continuously monitors the RSS feed of The Pirate Bay portal. For each new published torrent the RSS Parser gathers the  content infohash, from either the .torrent file or the magnet link\footnote{From 1st of March 2012, our tool will use exclusively magnet links for this purpose, as the Pirate Bay will stop serving .torrent files from that date.}, and also the publisher's username. Furthermore, the RSS Parser sends requests to the Tracker Connector.

\item \emph{Tracker Connector}: this module is responsible for connecting to the tracker for every torrent obtained by the RSS Parser. The main objective of the Tracker Connector is to obtain the IP address of the initial seeder. In those swarms where the  list of IP addresses returned by the tracker contains more peers than just one seeder, this module connects to all the peers and retrieves their bitfield in order to identify which one is the initial seeder.  If the IP address of the initial seeder matches with one of those included in the blacklist of fake IP addresses, this torrent is marked as fake.
 
\item \emph{The PirateBay Parser}: this module periodically connects to the Pirate Bay webpage associated to the different discovered publishers. Eventually, when a publisher's webpage (i.e. account) is removed from The Pirate Bay, the Pirate Bay Parser marks this username as fake.

\item \emph{Database}: It stores all the relevant information for the detection and evaluation of TorrentGuard. For each inspected torrent it stores detailed information such as the publisher's username and the initial seeder IP address (in case this is possible to obtain). More importantly, it includes two blacklists. The first one contains the infohashes of all the discovered fake torrents whereas the second one includes the IP addresses of fake publishers found so far.

\item \emph{Website Interface} and \emph{Vuze plugin}: The TorrentGuard functionality is publicly available throughout two different interfaces: a website\footnote{This application is available at \url{http://torrentguard.netcom.it.uc3m.es/}} and a Vuze plugin. These interfaces provide access to the blacklist of fake torrents allowing a user to verify if a torrent file is associated to a fake content before starting the download process.

\end{itemize}

Next, we describe the functionality of the integrated TorrentGuard tool detailing the interaction between the different modules as well as the configuration parameters. It uses The Pirate Bay portal in order to identify new fake publishers and the IP addresses from where they operate. Towards this end, the RSS Parser continuously monitors the RSS feed of The Pirate Bay portal to learn about new torrents and identify for each torrent the publisher's username. Furthermore, it sends a query to the Tracker Connector that retrieves the IP address of the initial seeder (if it is possible). Both, the publisher's username and IP address (i.e. IP address of the initial seeder) are stored in the database. In parallel, the Pirate Bay Parser periodically connects to the webpage of the different discovered publishers within The Pirate Bay. If we find that a publisher's account is removed, this user and all its torrents are marked as fake. In addition, we annotate this publisher's IP address as \emph{potential fake IP address}. If three different accounts associated to a given publisher's IP address are removed from The Pirate Bay, we consider that IP as a \emph{fake IP address}. From this moment on, any content published from that IP address is identified just after its birth and reported as fake. The number of removed accounts needed to mark an IP address as \emph{fake} is a configurable parameter in TorrentGuard. We decided to set up this parameter equal to three because as we will demonstrate in Section \ref{subsec:performance} this provides a negligible ratio of false positive and false negatives. Decreasing this value makes TorrentGuard more aggressive and may increase the number of false negatives. Rather increasing it  makes TorrentGuard more conservative what may increase the number of false positives.

Therefore, in the worst case, i.e. for new fake publishers, TorrentGuard employs the same time as The Pirate Bay to identify fake content. However, once the fake publisher's IP address has been identified, TorrentGuard is able to report fake content immediately after its publication. This provides a significant improvement compared to standard detection mechanisms. In other words, with TorrentGuard it is not necessary to manually report each fake user account as the existing solutions require. 

Furthermore, the current existing solutions are limited to the portal where they operate. For instance, in the case of The Pirate Bay, once a content is identified as fake it is removed from the portal but not from the BitTorrent Ecosystem. Rather, TorrentGuard is a cross-portal solution, that is able to identify the infohash of the fake content preventing its download independently of the source from where the user obtained the .torrent file: any BitTorrent portal or the DHT service.

\emph{In short, TorrentGuard is a novel tool that: $(i)$ reduces fake content detection time since it uses IP-based detection instead of username-based detection and $(ii)$ allows to identify a fake content in the whole BitTorrent ecosystem instead of in a single portal because it identifies the fake content using the infohash (an unique identifier of the content in the whole BitTorrent ecosystem) rather than the torrent-id of an specific portal}.

\subsection{TorrentGuard Performance}
\label{subsec:performance}

\begin{figure}[t]
		\centering
		\includegraphics[width=2.4in]{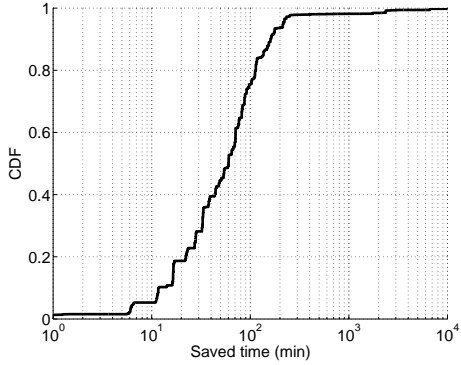}

		\caption{CDF of the saved time in fake content detection when using TorrentGuard in front of The Pirate Bay}
		\label{fig:cdf_time}
		\vspace{-0.6cm}

\end{figure}

We have evaluated the performance of TorrentGuard and compared it with the fake content detection mechanism used by The Pirate Bay during a testing period of 14 days. First, we count how many fake content published in The Pirate Bay are identified by the TorrentGuard just after its birth. Furthermore, we measure how long The Pirate Bay takes to identify these fake content. The obtained results show that TorrentGuard is able to early detect around 50\% of the fake content uploaded to The Pirate Bay. Moreover, Figure \ref{fig:cdf_time} represents the CDF of the time difference between the  detection instant of TorrentGuard and The Pirate Bay for these content. We observe, that TorrentGuard reduces the detection time 60 minutes in median. 
 Moreover, the reduction in detection time is higher than 2 hours for 20\% of the fake contents, and for some cases it goes up to several days.

Although previous results already demonstrate the significant improvement provided by our tool compared to the state of the art solution, the final objective of TorrentGuard is reducing the number of download events associated with fake content, thus preventing BitTorrent users facing malware and scam. Then, if TorrentGuard was widely used, it would have prevented almost 390K fake content downloads just during the 14 days of the evaluation period compared to The Pirate Bay. By extending this value to a complete year, we can state that TorrentGuard would be able to eliminate more than 10 millions fake content downloads per year compared to the existing The Pirate Bay solution. However, as stated before The Pirate Bay solution is specific for this portal but it is not applicable to the whole BitTorrent ecosystem. Specifically, in our dataset we identify around 950K fake content downloads occurring after The Pirate Bay identifies these content as fake. Rather, our proposed solution would be able to avoid also these downloads. Overall, TorrentGuard could avoid more than 1.35 millions fake content downloads in a period of two weeks. This means more than 35 millions in the course of a year. Finally, it is worth to mention that even this impressive number is only a lower bound since in our evaluation we only consider download events associated to few of the most important BitTorrent Trackers\footnote{For instance, \url{http://openbittorrent.com/}, \url{http://publicbt.com/} that are the two major Trackers in the BitTorrent ecosystem} but we do not consider download events coming from minor BitTorrent Trackers or the BitTorrent-associated DHT systems.

\emph{In a nutshell, our initial evaluation suggests that TorrentGuard could avoid up to tens of millions fake downloads per year. More importantly, this supposes (depending on the success of the fake publishers' strategies) up to hundreds of thousands of computer infections and scam episodes. Hence, our evaluation shows very promising results to incentive the BitTorrent community to use the TorrentGuard.}


\subsection{TorrentGuard Efficiency}

The efficiency of a detection system is typically characterised by the rate of false negative and false positive occurrences. In the specific case of TorrentGuard false negatives are represented by those fake torrents escaping our detection tool whereas false positives refer to those content classified as fake which actually are non fake ones. 

Exhaustively measuring the false negative rate is not scalable in the case of TorrentGuard since it would require to download and manually inspect every single content classified as legacy (i.e. non fake) by TorrentGuard. This means up to dozens of thousands of content every month. Instead, we have performed an affordable evaluation by downloading few dozens of torrents classified as legacy by TorrentGuard and manually inspecting them. We did not find any fake torrent among them. We can state, however, that our tool discovers all fake contents which are also detected by The Pirate Bay.

In order to evaluate the false positives rate of TorrentGuard, we focus on those Pirate Bay usernames whose account has not been deleted from The Pirate Bay but their content have been classified by TorrentGuard as fake. The first intuition is that TorrentGuard may be mistaken for some of these usernames. We have downloaded content from each of these referred Pirate Bay accounts and we did not find any non-fake content among them. Thus these content belong to fake publishers that have still not been detected by The Pirate Bay. 

\emph{In a nutshell, the performed evaluation suggests that TorrentGuard present a negligible rate of both false positive and false negative events.}

\subsection{Low impact of TorrentGuard External Dependencies}

In this subsection we discuss the external dependencies of TorrentGuard and demonstrate that they represent a minor limitation for the system.

\subsubsection{Dependency in The Pirate Bay}
We have explained above that TorrentGuard bases its operation in The Pirate Bay portal. We selected The Pirate Bay because it is the most important portal and one of the key elements of the BitTorrent ecosystem \cite{Zhang10}. Fake publishers could use other portals in order to not be detected by TorrentGuard, but then their visibility would be significantly affected. As future work, we plan to extend TorrentGuard to other portals. The requirements for these portals are: $(i)$ having a service to announce new published torrents (e.g. RSS or a webpage) and $(ii)$ having a system to report fake publishers (e.g. removing their accounts as in The Pirate Bay or marking fake content with special flags). It is worth to mention, that these two requirements are pretty standard and widely offered by the most significant BitTorrent portals such as Mininova \cite{mininova} or IsoHunt \cite{ISOHUNT}.

\subsubsection{Dependency on Users' Reports}
To the best of the authors' knowledge none existing software has the capacity of identifying a fake content under this context, i.e. the software should discern if the content is fake or not using as input the title of the content. For this purpose, we require the intervention of a human being. Hence, in practice we need at least one user's report to identify a fake content and its associated fake publisher. As discussed earlier, TorrentGuard can be configured to mark a fake publisher's IP address after the first user report (i.e. removed fake username account). However as stated before we prefer be more conservative and mark the IP as fake after 3 reports to minimize the false negatives.

\vspace{0.2cm}

\emph{In summary, the previous discussion demonstrates that the external dependencies of TorrentGuard do not affect seriously its performance. On the one hand, the dependency of TorrentGuard in a single portal can be overcome by extending the operation of TorrentGuard to multiple portals. It is worth to mention that the effectiveness of TorrentGuard will be directly related to the significance of the associated portals. On the other hand, the dependency on users' reports is inherent to any fake content detection system and cannot be removed until new semantic-enhanced software is implemented. Hence, the best we can do is minimize the dependency in users' reports and TorrentGuard achieves this objective.}

\subsection{Limitations of potential countermeasures to TorrentGuard}
\label{sec:countermeasures}

If TorrentGuard becomes widely used, it is likely that the fake publishers will react by defining new strategies (i.e. \emph{countermeasures}) that allow them to escape the control of TorrentGuard. Our tool identifies the fake publisher based on the IP address that it uses to publish the fake content. Hence, the fake publishers can use two reactive strategies. First, they can try to hide their IP address and second, they can try to perform their activity from a large number of IP addresses. In this subsection, we will discuss these strategies and their potential effectiveness.

\subsubsection{Hiding the Fake Publisher's IP address}

The most straightforward way to hide an IP address is the utilization of a proxy. In this case TorrentGuard will interpret that the fake activity is being performed from the proxy IP address and will banned this one. Hence this technique is not efficient against TorrentGuard.

The next option would be to consider a network of proxies such that the fake Publisher can use different proxies for publishing different fake content. This type of anonymisation services exist in the current Internet and are commonly used by regular BitTorrent users to hide its IP address during the process of illegal content downloads and TOR is an example \cite{tor}. In TOR, traffic from a source (a fake publisher in our case) is bounced through several relays until it reaches the destination. Hence, the destination see that packets are coming from the IP address of the last (or \emph{egress}) proxy and the IP address of the source cannot be identified. Furthermore, the egress proxy changes from one communication to another. Fake publishers could exploit the functionality of TOR to avoid its IP address being detected by TorrentGuard. TorrentGuard would then mark the IP addresses of TOR egress proxies as fake. Hence, if some non-fake publishers would use TOR, TorrentGuard would also mark their content as fake, thus increasing the false positives rate.

However, it is important to highlight that these anonymity services were not designed for supporting heavy traffic applications such as BitTorrent so that the performance offered to these services is typically poor. Indeed, TOR developers specifically state that TOR does not perform well with BitTorrent and is not designed for handling that type of traffic \cite{tor2}. To evaluate the performance degradation that a fake publisher would experiment using TOR we have run a very simple test that compare the performance of a regular BitTorrent download vs a download done with usage of TOR. For this purpose we have chosen a mid-popular torrent from The Pirate Bay (around 200 seeders and 300 lechers, 350,5 MB) and downloaded it 10 times with and without TOR usage. We have run the experiment in premises of our University (with a symmetric connection of 100  Mbps) and using a home ADSL (with a download and upload bandwidth of 6 Mbps and 320 kbps respectively). The results are presented in Tab. \ref{tab:tor}. They suggest that operating BitTorrent over TOR reduces the performance around 3 times independently of the speed of the access link. Therefore, the utilization of anonymisation networks by fake publishers would severely impact the performance (i.e. content download time) of the swarms associated to fake content. This would result in attracting a lower number of victims that would prefer faster downloads. In addition, we have revealed in Section \ref{sec:malicious} that the top fake publishers perform their activity from high speed services. This suggests that performance is a key aspect for their activity, thus anonymisation services seem to be a not appropriate option for them.

In summary, current solutions that could be used by a fake publisher in order to hide its IP address are either not efficient (e.g. single proxy) or incur an important performance degradation that seems to not be adequate for the fake publishers' activity.





\begin{table}

\center

\begin{tabular}{||p{0.2\textwidth}|| p{0.1\textwidth} | p{0.1\textwidth}||}
\hline \hline Type of connection &  Average Time & Average speed\\ 
\hline \hline 
University	&	6m 46s & 6.9 Mbit/s \\
University (with TOR)	&	20m 31s & 2.27 Mbit/s \\
Home ADSL	&	9m 59s &  4.68 Mbit/s\\
Home ADSL (with TOR)	&	31m 15s & 1.49 Mbit/s \\
\hline \hline 

\end{tabular} 

\caption{Average speed and download time of the file using BitTorrent with and without TOR}
\label{tab:tor}
\vspace{-1.2cm}
\end{table}

\subsubsection{Using multiple IP addresses}

The second countermeasure that a fake publisher could opt for is using a large number of IP addresses such that it always have undetected IP addresses to use for publishing fake content.
Next, we estimate the number of IP addresses that a fake publisher would need to perform its activity in the presence of our tool. TorrentGuard identifies an IP address as fake after detecting 3 fake user accounts in The Pirate Bay. Thus, TorrentGuard marks a content as fake starting from the 4$^{th}$ account used by the publisher. We demonstrated in Section \ref{sec:malicious} that top 5\% of fake publishers use in average 4 user accounts per day. Hence, a top fake publisher would need roughly 1 IP address per day in order to perform its activity and avoiding being blocked by TorrentGuard. In addition, we have seen that the activity of these publisher is performed from high speed servers located in data centres. Hence, these users would need to have access to around 30 IP addresses associated to high speed access links per month. 

In short, this strategy represents a double serious challenge: first, the fake publisher should be able to get continuously 30 new IP addresses per month and second, these IP addresses needs to be associated to high speed access links. This is rather difficult for regular Internet users and companies.


\vspace{0.2cm}

\emph{We can conclude that the studied countermeasures against TorrentGuard are either inefficient or unrealistic. Hence, the wide usage of TorrentGuard may lead to discourage fake publishers to perform their activity.}


\subsection{Torrent Guard Future Deployment}

In the previous subsections we have demonstrated the enormous potential of our TorrentGuard prototype. However, we believe that there is still room for improvement if BitTorrent portals and Trackers get involved in a next stage for the development of TorrentGuard. In this case, TorrentGuard could be extended to be a distributed platform in which trackers would identify the IP address of the initial seeder for every content and  BitTorrent portals would identify the infohash of fake torrents. BitTorrent portals would provide the infohash of fake torrents to trackers so that these would be able to blacklist the IP address associated to fake publishers and eliminate their associated swarms. Furthermore, trackers would report back to portals the infohash of every new fake torrent published from a blacklisted IP address so that portals can immediately remove the associated .torrent file.
The described system could store the information in a central server that interacts with both portals and trackers and maintain a central repository that can be accessed by users as well. Another option is running a complete distributed system in which trackers and portals exchange the information without the necessity of any central server. We believe that the involvement of major BitTorrent Portals and Trackers in this project would lead to reduce the presence of fake content to negligible levels\footnote{The authors of this paper have started a process to contact different Trackers and Portals to sense their interest in participating in the deployment of the described project.}.

\section{Related Work}
\label{sec:related_work}
\subsection{BitTorrent Measurement}
Several authors have used real data collection in order to understand different aspects of BitTorrent \cite{INFOCOM}\cite{Guo05}\cite{Piatek07}. Different methods of measuring the BitTorrent are described in \cite{comm}. However, only few works have looked at the content publishers \cite{Legout10}\cite{Zhang10}. The most extensive study of characterisation of BitTorrent ecosystem is presented in \cite{Zhang10}. This work includes discussion about BitTorrent publishers, defined by its username. We demonstrate in this paper that fake publishers cannot be identified by its username, instead they are identified by its IP address. The presence of the fake publishers was firstly mentioned in our previous work \cite{CONEXT}. Based on our initial observation, in this paper we perform a thorough analysis of fake publishers and their published content revealing their target, incentives and strategies and propose a novel solution to prevent users from downloading fake content.

\subsection{Fake content}
There are several studies presenting the possible threats in the Internet. In \cite{Zhaosheng08} authors state that 40\% of all computers are infected by botnets and can be controlled by attackers. Another study \cite{Moshchuk06} reports high presence of malware and spyware content in the Internet. Few previous works have studied the malware propagation through P2P systems \cite{Kalafut06}\cite{Shin06}\cite{Zhou05}. Specifically,  Kalafut et al. \cite{Kalafut06} analyse LimeWire whereas Shin et al. \cite{Shin06} analysed KaZaa. These authors look at the problem from the content perspective instead of the fake publisher perspective used in this paper. This avoids that they discover more sophisticated strategies as those reported in our study in which the content  is not the malware itself but includes a link to the malware. Similar content-based approach is applied in FakeDetector program \cite{FakeDetector} that looks for fake hashes in DirectConnect hubs (central servers to which downloaders connect) and reports found fake content to users and hub administrators. Finally, the authors of \cite{Kalafut06} propose to filter those content with a specific size since most of the malware content has specifically this size. Unfortunately, this solution is not valid for BitTorrent. Instead, we propose a more sophisticated solution (TorrentGuard) that provides early detection of fake content.

\section{Conclusions}
\label{sec:conclusions}

This paper presents the first comprehensive study about fake content in the BitTorrent ecosystem. For this purpose we use real data collected during a large-scale measurement study. The obtained results demonstrate that 35\% of all the content is fake. Moreover, just a few tens of users are responsible for most of the published fake content. Furthermore, more than 99\% of the fake torrents are associated with either malware or scam websites. This represents a serious threat for the BitTorrent ecosystem that must be eliminated or at least mitigated. Towards this end, we have implemented TorrentGuard, a novel tool for early detection of fake content. Based on our initial evaluation the widely usage of this tool may prevent the download of millions of fake content every year, thus contributing to reduce the number of computer infections and scam episodes faced by BitTorrent users.

\bibliographystyle{plain}

\bibliography{p2p.bib}

\end{document}